\documentclass[%
 reprint,
 amsmath,amssymb,
 aps,
]{revtex4-2}
\usepackage{times}
\usepackage{graphicx}
\usepackage{dcolumn}
\usepackage{bm}
\usepackage{xcolor}
\usepackage{braket}
\usepackage{hyperref}
\usepackage{mathtools}
\usepackage{orcidlink}
\usepackage{svg}
\allowdisplaybreaks
\begin{document}

\preprint{APS/123-QED}

\title{A Collective-Spin Derivation of the Uniform Magnon Hamiltonian in Cavity Magnonics}

\author{T. Aguiar%
 \orcidlink{0000-0002-1141-2241}}
\email{taguiarcf@gmail.com, corresponding author.}
\affiliation{Instituto de F\'\i sica  Gleb Wataghin, Universidade Estadual de Campinas, 13083-859, Campinas, SP, Brazil} 
\author{M. C. de Oliveira%
 \orcidlink{0000-0003-2251-2632}}\email{marcos@ifi.unicamp.br}
\affiliation{Instituto de F\'\i sica  Gleb Wataghin, Universidade Estadual de Campinas, 13083-859, Campinas, SP, Brazil}%

\date{\today}

\begin{abstract}
We present a direct collective-spin derivation of the effective
uniform-mode Hamiltonian used in cavity magnonics. Starting from a
nearest-neighbor Heisenberg ferromagnet coupled to long-wavelength
magnetic fields, we show that the relevant dynamics can be restricted
to the fully symmetric spin sector, where the exchange interaction
contributes only a constant energy shift and the ferromagnet behaves
as a macrospin of length \(Ns\). Applying the Holstein--Primakoff
transformation directly to this total spin yields the usual uniform
magnon mode and its leading nonlinear corrections without first
introducing site-resolved bosonic operators. This collective
formulation makes explicit the interpretation of the ferromagnet as a
synthetic large-spin atom and provides a compact route to the effective
Hamiltonians used in driven and Floquet cavity magnonics. As a physical
consequence, the leading nonlinear correction produces an
occupation-dependent reduction of the effective magnon--photon
coupling, providing a simple signature of finite-spin saturation under
strong uniform-mode driving.
\end{abstract}

\maketitle


\section{\label{sec:level1}Introduction}

The field of quantum magnonics has entered a new stage with the preparation and control of coherent hybrid magnon--polariton states under Floquet dynamics~\cite{xu2020floquet,xu2023dynamical,pishehvar2025resonance}. 
By periodically driving a ferromagnet with an oscillatory magnetic field aligned with the static bias field, one can modulate the precession of the magnetization. 
This time-periodic modulation enables the observation of nontrivial phenomena, such as Floquet sidebands, \textit{Rabi oscillations}, and \textit{Autler--Townes} splittings, closely analogous to effects well known in atomic and molecular physics.
Perturbative input--output approaches have also been developed to describe
Floquet cavity-magnonics spectra and magnon energy shifts in this regime~\cite{Aguiar2026}.

In light of these developments, it is useful to revisit the fundamental physical system underlying cavity-magnonics experiments: a ferromagnetic sample coupled to controlled electromagnetic fields. 
The relevant fields are: (i) a static bias field, which defines the magnetic ground state and sets the natural precession frequency; (ii) an in-plane field perpendicular to the static field, which drives the magnetization precession; and (iii) an out-of-plane, or Floquet field, parallel to the static field, which modulates the precession dynamics. 
Through precise control of these fields at cryogenic temperatures, it is possible to prepare and manipulate robust quantum magnon states. 
This has enabled the resolution of individual spin flips within a macroscopic ferromagnet~\cite{lachance2017resolving,lachance2020entanglement} and the coherent manipulation of superpositions of distinct magnon states~\cite{xu2023quantum}. 
These achievements highlight cavity magnonics as a versatile platform for quantum information processing, hybrid quantum technologies, and fundamental studies of light--matter interactions.

The theoretical framework most commonly used to describe magnonic systems starts from a site-resolved second quantization of the Heisenberg Hamiltonian, followed by the Holstein--Primakoff (HP) transformation~\cite{prabhakar2009spin} applied to each atomic spin. 
This procedure is powerful and general, since it naturally produces the full spin-wave spectrum. 
However, in the long-wavelength regime relevant to many cavity-magnonics experiments, the dominant excitation is the uniform, magnetostatic mode. 
In this case, the site-resolved route may obscure the collective nature of the excitation and make the derivation of nonlinear corrections unnecessarily indirect.

To set the notation, we recall the conventional site-resolved HP transformation,
\begin{eqnarray}
\hat{\text{s}}_{\boldsymbol{\mathfrak{i}}}^+ 
&\rightarrow& 
(2s)^{1/2}
\left(
\mathbb{I}
-
\dfrac{
\hat{\text{m}}_{\boldsymbol{\mathfrak{i}}}^{\dagger}
\hat{\text{m}}_{\boldsymbol{\mathfrak{i}}}
}{2s}
\right)^{1/2}
\hat{\text{m}}_{\boldsymbol{\mathfrak{i}}},
\label{HP_transf1}
\\
\hat{\text{s}}_{\boldsymbol{\mathfrak{i}}}^- 
&\rightarrow& 
(2s)^{1/2}
\hat{\text{m}}_{\boldsymbol{\mathfrak{i}}}^{\dagger}
\left(
\mathbb{I}
-
\dfrac{
\hat{\text{m}}_{\boldsymbol{\mathfrak{i}}}^{\dagger}
\hat{\text{m}}_{\boldsymbol{\mathfrak{i}}}
}{2s}
\right)^{1/2},
\\
\hat{\text{s}}_{\boldsymbol{\mathfrak{i}}}^z 
&\rightarrow& 
s-
\hat{\text{m}}_{\boldsymbol{\mathfrak{i}}}^{\dagger}
\hat{\text{m}}_{\boldsymbol{\mathfrak{i}}},
\label{HP_transf3}
\end{eqnarray}
with
\[
\left[
\hat{\text{m}}_{\boldsymbol{\mathfrak{i}}},
\hat{\text{m}}_{\boldsymbol{\mathfrak{i}}'}^{\dagger}
\right]
=
\delta_{\boldsymbol{\mathfrak{i}},\boldsymbol{\mathfrak{i}}'}
\mathbb{I}.
\]
Here, $\hat{\text{s}}_{\boldsymbol{\mathfrak{i}}}^{\pm,z}$ are the spin operators at site $\boldsymbol{\mathfrak{i}}$, while $\hat{\text{m}}_{\boldsymbol{\mathfrak{i}}}^{\dagger}$ and $\hat{\text{m}}_{\boldsymbol{\mathfrak{i}}}$ create and annihilate, respectively, a local spin flip at that site. 
With the convention adopted here, the fully polarized state along the positive $z$ direction is the bosonic vacuum, and a magnon corresponds to a reduction of the spin projection along $z$.

For small magnon populations, one expands the square roots in Eqs.~\eqref{HP_transf1}--\eqref{HP_transf3} and transforms the bosonic operators to reciprocal space. 
This leads to a description of the ferromagnet in terms of spin-wave modes, each behaving, to lowest order, as a quantum harmonic oscillator with an energy determined by the corresponding dispersion relation. 
In cavity-magnonics experiments, one typically retains only the uniform mode, while the remaining spin-wave modes are neglected.

In this work, we present a direct collective-spin derivation of the effective uniform-mode Hamiltonian used in cavity magnonics. 
Starting again from the Heisenberg Hamiltonian, we recast the problem in terms of the total spin of the ferromagnet and apply the HP transformation directly at this collective level. 
This route bypasses the intermediate site-resolved bosonic representation and makes explicit the interpretation of the ferromagnet as a macrospin, or synthetic atom, of total spin $Ns$. 
The resulting expansion is controlled by the condition
\[
\langle \hat{\text{M}}^{\dagger}\hat{\text{M}}\rangle \ll 2Ns,
\]
where $\hat{\text{M}}^{\dagger}$ creates an excitation of the collective uniform mode. 
Thus, the small parameter is naturally associated with the density of collective spin flips relative to the total spin magnitude.

Our approach is based on three assumptions commonly satisfied in cavity-magnonics experiments:
(i) the ferromagnetic sample is positioned at a maximum of the standing-wave cavity mode;
(ii) the cavity wavelength, typically of order $1\,\mathrm{cm}$, is much larger than the sample size, typically of order $100\,\mu\mathrm{m}$, ensuring an approximately uniform excitation; and
(iii) the system is prepared close to its magnetic ground state, with all $N$ atomic spins aligned along the static bias field.

Under these conditions, the effective Hamiltonian expressed in terms of total spin operators captures the dominant uniform mode and exposes the leading nonlinear corrections in a transparent perturbative hierarchy. 
The formulation is equivalent to the conventional site-resolved derivation followed by projection onto the uniform mode, but it emphasizes the collective character of the excitation from the outset and provides a compact route to the nonlinear magnon--photon interactions relevant under strong driving and Floquet modulation.

This paper is organized as follows. 
In Sec.~II, we introduce the spin Hamiltonian relevant to cavity-magnonics experiments and derive the equations of motion for the reciprocal-space spin modes. 
In Sec.~III, we apply the Holstein--Primakoff transformation to the total spin and construct the effective Hamiltonian in terms of collective excitations. 
In Sec.~IV, we derive the finite-spin correction to the magnon--photon coupling and show that the leading nonlinear term produces an occupation-dependent reduction of the effective coupling under strong uniform-mode driving. 
In Sec.~V, we compare the conventional site-resolved route with the collective-spin derivation and discuss the physical interpretation of the ferromagnet as a synthetic large-spin atom. 
Finally, in Sec.~VI, we summarize our conclusions.

\section{\label{sec:spin_hamiltonian}Spin Hamiltonian in Cavity Magnonics Experiments}

We consider a macroscopic ferromagnet with cubic geometry. 
This choice is made for simplicity, since it allows one to impose simple boundary conditions and to write the spin modes in reciprocal space in a compact form. 
The crystal is modeled as a cubic lattice of \(N\) spin-\(s\) atoms with lattice parameter \(a\). 
Surface effects associated with the actual sample geometry, such as demagnetization fields, are neglected. 
We also assume that the temperature is sufficiently low for thermal excitations to be negligible. 
Thus, by applying a static field \(B_0\) along the \(z\) direction, the ferromagnet is prepared close to its magnetic ground state, in which all spin sites are aligned with the static field.\footnote{The indexation conventions used in all equations follow those of Dyson's seminal article on spin waves~\cite{dyson1956general}.}

Keeping only nearest-neighbor exchange interactions, the unperturbed spin Hamiltonian is
\begin{eqnarray}
     \hat{H}_S^{(0)}
     &=&
     -J
     \sum_{\boldsymbol{\mathfrak{i}},\boldsymbol{\mathfrak{j}}}
     \hat{\boldsymbol{\text{s}}}_{\boldsymbol{\mathfrak{i}}}
     \cdot
     \hat{\boldsymbol{\text{s}}}_{\boldsymbol{\mathfrak{i}}+\boldsymbol{\mathfrak{j}}}
     -
     \gamma B_0 
     \sum_{\boldsymbol{\mathfrak{i}}}
     \hat{\text{s}}_{\boldsymbol{\mathfrak{i}}}^z ,
     \label{bare_spin_hamiltonian}
\end{eqnarray}
where $\gamma=28.02~\text{GHz}/\text{T}$ is the gyromagnetic ratio of the electron and
\begin{eqnarray}
    \sum_{\boldsymbol{\mathfrak{i}},\boldsymbol{\mathfrak{j}}}
    \hat{\boldsymbol{\text{s}}}_{\boldsymbol{\mathfrak{i}}}
    \cdot
    \hat{\boldsymbol{\text{s}}}_{\boldsymbol{\mathfrak{i}}+\boldsymbol{\mathfrak{j}}}
    =
    \sum_{\boldsymbol{\mathfrak{i}},\boldsymbol{\mathfrak{j}}}
    \left[
    \dfrac{1}{2}
    \left(
    \hat{\text{s}}_{\boldsymbol{\mathfrak{i}}}^+
    \hat{\text{s}}_{\boldsymbol{\mathfrak{i}}+\boldsymbol{\mathfrak{j}}}^-
    +
    \hat{\text{s}}_{\boldsymbol{\mathfrak{i}}}^-
    \hat{\text{s}}_{\boldsymbol{\mathfrak{i}}+\boldsymbol{\mathfrak{j}}}^+
    \right)
    +
    \hat{\text{s}}_{\boldsymbol{\mathfrak{i}}}^z
    \hat{\text{s}}_{\boldsymbol{\mathfrak{i}}+\boldsymbol{\mathfrak{j}}}^z
    \right].
    \label{exchange_decomposition}
\end{eqnarray}
Here, 
\(\boldsymbol{\mathfrak{i}}=(\mathfrak{i}_x,\mathfrak{i}_y,\mathfrak{i}_z)\) labels the atomic sites of the cubic lattice, and 
\(\boldsymbol{\mathfrak{j}}=(\mathfrak{j}_x,\mathfrak{j}_y,\mathfrak{j}_z)\) labels the relative position of the nearest neighbors of each site. 
For a cubic lattice,
\[
\boldsymbol{\mathfrak{j}}
\in
\left\{
(\pm1,0,0),(0,\pm1,0),(0,0,\pm1)
\right\}.
\]
We take \(n^3=N\), where \(n\) is the number of sites along each spatial direction. 
The parameter \(J\) denotes the nearest-neighbor exchange coupling, written in frequency units. 
Multiplying the dimensionless lattice indices by the lattice parameter \(a\) recovers the corresponding position vectors in physical space.

\begin{figure}[ht]
    \centering
    \includegraphics[width=1\linewidth]{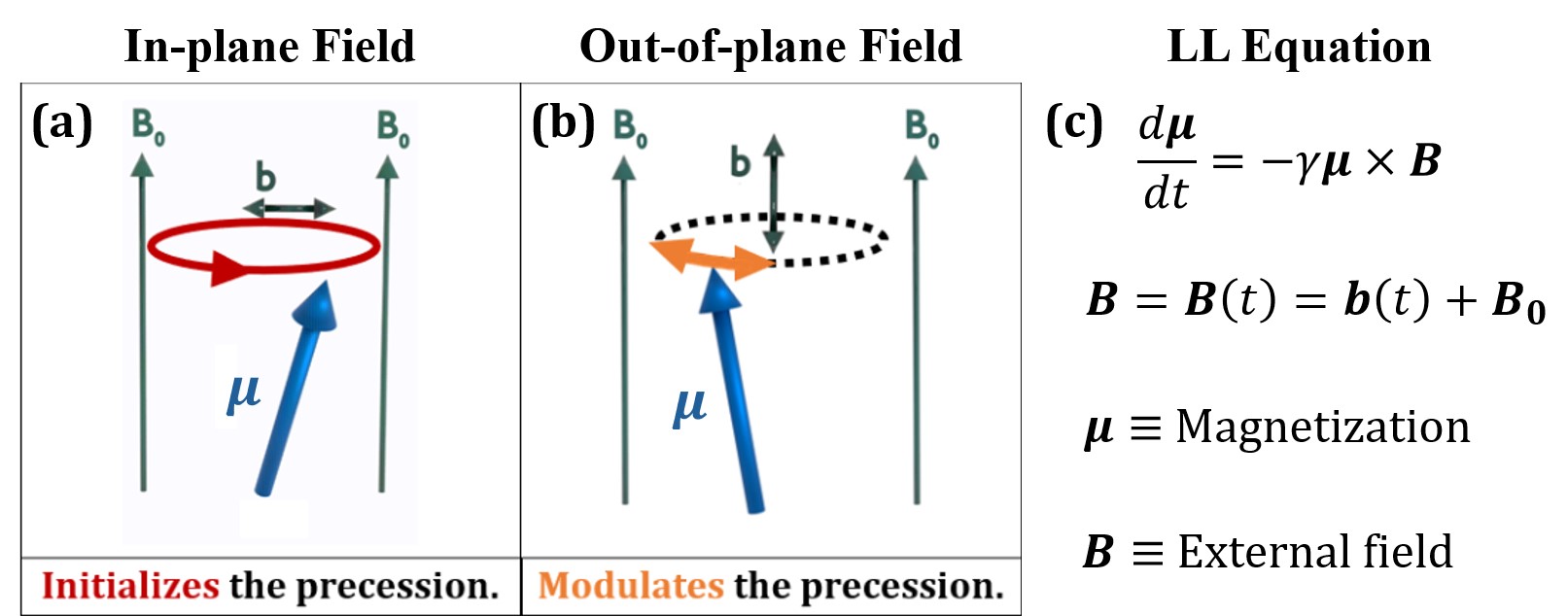}
    \caption{Magnetization dynamics induced by different excitation fields. (a) In-plane field dynamics. (b) Out-of-plane field dynamics. (c) Landau-Lifshitz equation, where $\boldsymbol{\mu}$ is the magnetization vector and $\boldsymbol{B}(t)$ is the total magnetic field, composed of an oscillatory field component, $\boldsymbol{b}(t)$, and a static field, $\boldsymbol{B}_0$.  Adapted from  \cite{nanomag_rep}.}
    \label{fig:types_prec}
\end{figure}

In microwave cavity setups, the ferromagnet is usually placed at a position of maximum magnetic flux density. 
Since the microwave wavelength is typically much larger than the sample size, the excitation is approximately uniform over the ferromagnetic volume. 
The applied fields considered here can be separated into two classes, according to the classical precession dynamics of the magnetization, as illustrated in Fig.~\ref{fig:types_prec}. 
An in-plane drive is perpendicular to the static field and induces the magnetization to acquire a component in the \(xy\) plane. 

This transverse component then precesses around the static field. 
In the simplified treatment adopted here, resonance is reached when the drive frequency matches
\[
g_0=\gamma B_0.
\]

An out-of-plane drive is parallel to the static field and modulates the precession frequency. 
Together with the precession, this produces a modulated magnetization dynamics described, at the classical level, by the Landau--Lifshitz equation (See Fig.\ref{fig:types_prec}).

In the quantum description, the magnetization dynamics, denoted by \(\boldsymbol{\mu}\), is obtained from the expectation value of the total spin angular momentum operator through
\[
\boldsymbol{\mu}=-\dfrac{\gamma\hbar \braket{\hat{\textbf{S}}}}{V},
\]
where $V$ is the volume of the magnetic material and \(\hat{\textbf{S}}\) denotes the total spin in physical space. 

The Hamiltonian in the presence of the in-plane and out-of-plane fields can then be written as
\begin{align}
    \hat{H}_S
    =&
    -J
    \sum_{\boldsymbol{\mathfrak{i}},\boldsymbol{\mathfrak{j}}}
    \hat{\boldsymbol{\text{s}}}_{\boldsymbol{\mathfrak{i}}}
    \cdot
    \hat{\boldsymbol{\text{s}}}_{\boldsymbol{\mathfrak{i}}+\boldsymbol{\mathfrak{j}}}
    -
    g_0
    \sum_{\boldsymbol{\mathfrak{i}}}
    \hat{\text{s}}_{\boldsymbol{\mathfrak{i}}}^z
    \nonumber\\
    &
    -
    g_c
    \sum_{\boldsymbol{\mathfrak{i}}}
    \hat{\text{s}}_{\boldsymbol{\mathfrak{i}}}^x
    \left(
    \hat{c}_{\boldsymbol{\lambda}_c}
    e^{i\boldsymbol{\lambda}_c\cdot\boldsymbol{\mathfrak{i}}}
    +
    \hat{c}_{\boldsymbol{\lambda}_c}^{\dagger}
    e^{-i\boldsymbol{\lambda}_c\cdot\boldsymbol{\mathfrak{i}}}
    \right)
    \nonumber\\
    &
    -
    g_b
    \sum_{\boldsymbol{\mathfrak{i}}}
    \hat{\text{s}}_{\boldsymbol{\mathfrak{i}}}^z
    \left(
    \hat{b}_{\boldsymbol{\lambda}_b}
    e^{i\boldsymbol{\lambda}_b\cdot\boldsymbol{\mathfrak{i}}}
    +
    \hat{b}_{\boldsymbol{\lambda}_b}^{\dagger}
    e^{-i\boldsymbol{\lambda}_b\cdot\boldsymbol{\mathfrak{i}}}
    \right).
    \label{full_spin_hamiltonian}
\end{align}
Here, \(\hat{c}_{\boldsymbol{\lambda}_c}\) and \(\hat{b}_{\boldsymbol{\lambda}_b}\) are the annihilation operators of the in-plane and out-of-plane field modes, respectively, and \(g_c\) and \(g_b\) are their corresponding single-spin coupling strengths. 
The reciprocal-space vectors are denoted by
\[
\boldsymbol{\lambda}=(\lambda_x,\lambda_y,\lambda_z),
\]
with
\[
\boldsymbol{\lambda}
\in
\left[
-\pi,
\left(\dfrac{n}{2}-1\right)\dfrac{2\pi}{n}
\right]^3,
\qquad
n^3=N.
\]

For fields whose wavelength is much larger than the size of the ferromagnet, the relevant field modes are approximately uniform, so that \(\boldsymbol{\lambda}_b,\boldsymbol{\lambda}_c\approx\boldsymbol{0}\). 
Introducing the reciprocal-space spin modes
\begin{align}
\hat{\boldsymbol{S}}_{\boldsymbol{\lambda}}
=
N^{-1/2}
\sum_{\boldsymbol{\mathfrak{i}}}
\hat{\boldsymbol{\text{s}}}_{\boldsymbol{\mathfrak{i}}}
e^{i\boldsymbol{\lambda}\cdot\boldsymbol{\mathfrak{i}}},
\label{spin_mode_definition}
\end{align}
and the exchange form factor
\begin{align}
u_{\boldsymbol{\lambda}}
=
\dfrac{1}{Z}
\sum_{\boldsymbol{\mathfrak{j}}}
e^{i\boldsymbol{\lambda}\cdot \boldsymbol{\mathfrak{j}}},
\label{form_factor_definition}
\end{align}
the Hamiltonian becomes
\begin{align}
   \hat{H}_S
   =
   &
   -JZ
   \sum_{\boldsymbol{\lambda}}
   u_{\boldsymbol{\lambda}}
   \hat{\boldsymbol{S}}_{\boldsymbol{\lambda}}
   \cdot
   \hat{\boldsymbol{S}}_{-\boldsymbol{\lambda}}
   -
   g_0 N^{1/2}\hat{S}_{\boldsymbol{0}}^z
   \nonumber\\
   &
   -
   \dfrac{g_c}{2} N^{1/2}
   \left(
   \hat{c}+\hat{c}^{\dagger}
   \right)
  \left(
\hat{S}_{\boldsymbol{0}}^+
+
\hat{S}_{\boldsymbol{0}}^-
\right)
   \nonumber\\
   &
   -
   g_b N^{1/2}
   \left(
   \hat{b}+\hat{b}^{\dagger}
   \right)
   \hat{S}_{\boldsymbol{0}}^z .
   \label{spin_ham_in_off}
\end{align}
where we used $\hat{S}_{\boldsymbol{0}}^x
=
\dfrac{1}{2}
\left(
\hat{S}_{\boldsymbol{0}}^+
+
\hat{S}_{\boldsymbol{0}}^-
\right)$. Here \(Z\) is the number of nearest neighbors, with \(Z=6\) for the cubic lattice, and we have defined the uniform field modes as
\[
\hat{c}_{\boldsymbol{0}}\equiv \hat{c},
\qquad
\hat{b}_{\boldsymbol{0}}\equiv \hat{b}.
\]
The form factor \(u_{\boldsymbol{\lambda}}\) encodes the nearest-neighbor structure of the lattice. 
Throughout this work, \(\hat{\mathbf{s}}_i\) denotes the spin operator at lattice site \(i\), while
\(\hat{\mathbf{S}}\equiv\sum_i\hat{\mathbf{s}}_i\) denotes the total spin. Using the definition of the reciprocal-space spin modes in Eq.\eqref{spin_mode_definition}, we find the expression for the uniform spin mode, 
\[
\hat{\boldsymbol{S}}_0=N^{-1/2}\hat{\mathbf{S}}.
\]

\subsection{\label{subsec:spin_eom_reciprocal}Spin Equations of Motion in Reciprocal Space}

The reciprocal-space spin operators satisfy the commutation relations~\cite{dyson1956general}
\begin{align} 
\left[
\hat{S}_{\boldsymbol{\lambda}}^{\pm},
\hat{S}_{\boldsymbol{\sigma}}^{z}
\right]
&=
\pm N^{-1/2}
\hat{S}_{\boldsymbol{\lambda}+\boldsymbol{\sigma}}^{\pm},
\label{comm_relations1}\\
\left[
\hat{S}_{\boldsymbol{\lambda}}^{+},
\hat{S}_{\boldsymbol{\sigma}}^{-}
\right]
&=
2 N^{-1/2}
\hat{S}_{\boldsymbol{\lambda}+\boldsymbol{\sigma}}^{z}.
\label{comm_relations}
\end{align}
Using these relations, together with the centrosymmetry of the reciprocal-space summation over \(\boldsymbol{\lambda}\), one obtains the Heisenberg equations of motion for the operators \(\hat{S}_{\boldsymbol{\sigma}}^{\pm,z}\). 
The full expressions for nonuniform spin modes are given in Appendix~\ref{eq_motion_spin}. 
They show that, in general, a spin mode \(\boldsymbol{\sigma}\) may couple to combinations of modes with indices \(\boldsymbol{\sigma}+\boldsymbol{\lambda}\).

The uniform mode, however, has a special role. 
For \(\boldsymbol{\sigma}=\boldsymbol{0}\), the commutation relations reduce to
\begin{align} 
\left[
\hat{S}_{\boldsymbol{\lambda}}^{\pm},
\hat{S}_{\boldsymbol{0}}^{z}
\right]
&=
\pm N^{-1/2}
\hat{S}_{\boldsymbol{\lambda}}^{\pm},
\\
\left[
\hat{S}_{\boldsymbol{\lambda}}^{+},
\hat{S}_{\boldsymbol{0}}^{-}
\right]
&=
2 N^{-1/2}
\hat{S}_{\boldsymbol{\lambda}}^{z}.
\label{comm_relations_uniform_2}
\end{align}
Because the ferromagnet is prepared close to its fully polarized ground state and the applied fields are uniform over the sample, the dominant dynamics is carried by the uniform spin mode. 
Moreover, the exchange contribution does not generate coupling of the uniform mode to nonuniform modes. 
The equations of motion for \(\hat{S}_{\boldsymbol{0}}^{\pm,z}\) therefore close among themselves:
\begin{align}
    \dfrac{d}{dt}\hat{S}_{\boldsymbol{0}}^{-}
    =&
    ig_0\hat{S}_{\boldsymbol{0}}^-
    +
    ig_b\hat{S}_{\boldsymbol{0}}^{-}
    \left(
    \hat{b}+\hat{b}^{\dagger}
    \right)
    \nonumber\\
    &
    -
    ig_c 
    \hat{S}_{\boldsymbol{0}}^z
    \left(
    \hat{c}+\hat{c}^{\dagger}
    \right),
    \label{eom_Sminus_uniform}
    \\
    \dfrac{d}{dt}\hat{S}_{\boldsymbol{0}}^{+}
    =&
    -ig_0\hat{S}_{\boldsymbol{0}}^+
    -
    ig_b\hat{S}_{\boldsymbol{0}}^{+}
    \left(
    \hat{b}+\hat{b}^{\dagger}
    \right)
    \nonumber\\
    &
    +
    ig_c 
    \hat{S}_{\boldsymbol{0}}^z
    \left(
    \hat{c}+\hat{c}^{\dagger}
    \right),
    \\
    \dfrac{d}{dt}\hat{S}_{\boldsymbol{0}}^{z}
    =&
    i\frac{g_c}{2}
    \left(
    \hat{S}_{\boldsymbol{0}}^+
    -
    \hat{S}_{\boldsymbol{0}}^-
    \right)
    \left(
    \hat{c}+\hat{c}^{\dagger}
    \right).
    \label{eom_Sz_uniform}
\end{align}

In the experimental regime considered here, the ferromagnet is initially prepared close to the fully polarized state, so that
\[
\braket{\hat{\text{S}}_z(0)}=Ns,
\]
or equivalently
\[
\braket{\hat{S}_{\boldsymbol{0}}^z(0)}=N^{1/2}s
\]
in the reciprocal-space representation. 
Thus, restricting the dynamics to the uniform sector, the effective spin Hamiltonian can be written as
\begin{align}
   \hat{H}_S^{\mathrm{eff}}
   =
   &
   -JZ
   \hat{\boldsymbol{S}}_{\boldsymbol{0}}
   \cdot
   \hat{\boldsymbol{S}}_{\boldsymbol{0}}
   -
   g_0 N^{1/2}
   \hat{S}_{\boldsymbol{0}}^z
   \nonumber\\
   &
   -
   \dfrac{g_c}{2} N^{1/2}
   \left(
   \hat{c}_{\boldsymbol{0}}
   +
   \hat{c}_{\boldsymbol{0}}^{\dagger}
   \right)
   \left(
\hat{S}_{\boldsymbol{0}}^+
+
\hat{S}_{\boldsymbol{0}}^-
\right)
   \nonumber\\
   &
   -
   g_b N^{1/2}
   \left(
   \hat{b}_{\boldsymbol{0}}
   +
   \hat{b}_{\boldsymbol{0}}^{\dagger}
   \right)
   \hat{S}_{\boldsymbol{0}}^z .
   \label{spin_ham_in_off_unif}
\end{align}
Since
\[
\hat{\boldsymbol{S}}_{\boldsymbol{0}}
=
N^{-1/2}
\hat{\textbf{S}},
\]
the exchange term in the uniform sector is proportional to the total-spin Casimir operator,
\begin{align}
\hat{\boldsymbol{S}}_{\boldsymbol{0}}
\cdot
\hat{\boldsymbol{S}}_{\boldsymbol{0}}
=
\dfrac{1}{N}
\hat{\textbf{S}}
\cdot
\hat{\textbf{S}}.
\end{align}
Within the fully symmetric spin sector, where the total spin is \(\text{S}=Ns\), this gives
\begin{align}
\hat{\boldsymbol{S}}_{\boldsymbol{0}}
\cdot
\hat{\boldsymbol{S}}_{\boldsymbol{0}}
=
s(Ns+1)\mathbb{I}.
\label{uniform_spin_casimir}
\end{align}
Therefore, in the collective uniform-mode description, the exchange interaction contributes only a constant energy shift. 
The remaining terms describe the coupling of the collective spin to the static in-plane and out-of-plane fields.

\section{\label{sec:collective_hp}The HP Transformation of the Total Spin}

The next step is to rewrite the effective Hamiltonian in terms of the total spin in physical space. 
Since
\[
\hat{\boldsymbol{S}}_{\boldsymbol{0}}
=
N^{-1/2}\hat{\textbf{S}},
\]
Eq.~\eqref{spin_ham_in_off_unif} becomes
\begin{align}
   \hat{H}_S^{\mathrm{eff}}
   =&
   -JZ s(Ns+1)\mathbb{I}
   -g_0 \hat{\text{S}}^z
   \nonumber\\
   &
   - \dfrac{g_c}{2}
   \left(
   \hat{c}+\hat{c}^{\dagger}
   \right)
   \left(
   \hat{\text{S}}^+
   +
   \hat{\text{S}}^-
   \right)
   \nonumber\\
   &
   - g_b
   \left(
   \hat{b}+\hat{b}^{\dagger}
   \right)
   \hat{\text{S}}^z ,
   \label{spin_ham_in_off_tot}
\end{align}
where we have defined
\[
\hat{c}\equiv \hat{c}_{\boldsymbol{0}},
\qquad
\hat{b}\equiv \hat{b}_{\boldsymbol{0}},
\]
for simplicity. The exchange contribution is a constant in the fully symmetric sector with total spin \(Ns\), and therefore does not affect the uniform-mode dynamics.

Equation~\eqref{spin_ham_in_off_tot} suggests a collective second-quantization scheme in which the Holstein--Primakoff transformation is applied directly to the total spin of the ferromagnet, rather than to each individual atomic spin. 
We choose the fully polarized state along the positive \(z\) direction as the bosonic vacuum. 
With this convention, a magnon corresponds to a reduction of the total spin projection by one unit, and the collective HP transformation reads
\begin{eqnarray}
   \hat{\text{S}}^+
   &\rightarrow&
   \left(2Ns\right)^{1/2}
   \left(
   \mathbb{I}
   -
   \frac{
   \hat{\text{M}}^{\dagger}\hat{\text{M}}
   }{2Ns}
   \right)^{1/2}
   \hat{\text{M}},
   \label{collective_HP_+}
   \\
   \hat{\text{S}}^-
   &\rightarrow&
   \left(2Ns\right)^{1/2}
   \hat{\text{M}}^{\dagger}
   \left(
   \mathbb{I}
   -
   \frac{
   \hat{\text{M}}^{\dagger}\hat{\text{M}}
   }{2Ns}
   \right)^{1/2},
   \\
   \hat{\text{S}}^z
   &\rightarrow&
   Ns-\hat{\text{M}}^{\dagger}\hat{\text{M}},
   \label{collective_HP_z}
\end{eqnarray}
with
\begin{equation}
\left[
\hat{\text{M}},
\hat{\text{M}}^{\dagger}
\right]
=
\mathbb{I}.
\label{collective_boson_commutator}
\end{equation}
Here, \(\hat{\text{M}}^\dagger\) creates a collective spin flip of the entire ferromagnet, i.e., an excitation of the uniform magnetic mode, and should not be confused with the magnetization, denoted here by \(\boldsymbol{\mu}\).

Substituting Eqs.~\eqref{collective_HP_+}--\eqref{collective_HP_z} into Eq.~\eqref{spin_ham_in_off_tot}, and omitting the constant exchange contribution, gives
\begin{align}
   \hat{H}_M
   =&
   -\dfrac{g_c}{2}
   \left(2Ns\right)^{1/2}
   \left[
   \hat{\text{M}}^{\dagger}
   \left(
   \mathbb{I}
   -
   \frac{
   \hat{\text{M}}^{\dagger}\hat{\text{M}}
   }{2Ns}
   \right)^{1/2}
   +
   \mathrm{h.c.}
   \right]
   \left(
   \hat{c}+\hat{c}^{\dagger}
   \right)
   \nonumber\\
   &
   -
   g_b
   \left(
   Ns-\hat{\text{M}}^{\dagger}\hat{\text{M}}
   \right)
   \left(
   \hat{b}+\hat{b}^{\dagger}
   \right)
   +
   g_0
   \hat{\text{M}}^{\dagger}\hat{\text{M}} .
   \label{mag_ham_exact_collective}
\end{align}
In the low-excitation regime,
\[
\langle \hat{\text{M}}^{\dagger}\hat{\text{M}}\rangle
\ll 2Ns,
\]
the square root can be expanded perturbatively. 
Keeping the leading nonlinear correction yields
\begin{align}
   \hat{H}_M
   \approx&
   -\dfrac{g_c}{2}
   \left(2Ns\right)^{1/2}
   \left(
   \hat{\text{M}}
   +
   \hat{\text{M}}^{\dagger}
   \right)
   \left(
   \hat{c}
   +
   \hat{c}^{\dagger}
   \right)
   \nonumber\\
   &
   -
   g_b
   \left(
   Ns-\hat{\text{M}}^{\dagger}\hat{\text{M}}
   \right)
   \left(
   \hat{b}
   +
   \hat{b}^{\dagger}
   \right)
   +
   g_0
   \hat{\text{M}}^{\dagger}\hat{\text{M}}
   \nonumber\\
   &
   +
   \frac{g_c}{4\left(2Ns\right)^{1/2}}
   \left(
   \hat{\text{M}}^{\dagger 2}\hat{\text{M}}
   +
   \hat{\text{M}}^{\dagger}\hat{\text{M}}^2
   \right)
   \left(
   \hat{c}
   +
   \hat{c}^{\dagger}
   \right).
   \label{mag_ham_magneto_simp}
\end{align}
This expression displays the usual linear coupling between the uniform magnon mode and the in-plane cavity field, the out-of-plane modulation of the magnon frequency, and the leading nonlinear correction arising from the collective HP expansion.

It is useful to relate the collective bosonic operator introduced above to the conventional uniform magnon operator. 
In the standard site-resolved treatment, the local HP bosons \(\hat{\text{m}}_{\boldsymbol{\mathfrak{i}}}\) are transformed to reciprocal space, and the uniform magnon operator is defined as
\begin{equation}
\hat{m}_{\boldsymbol{0}}
=
\frac{1}{\sqrt{N}}
\sum_{\boldsymbol{\mathfrak{i}}}
\hat{\text{m}}_{\boldsymbol{\mathfrak{i}}}.
\label{uniform_magnon_operator}
\end{equation}
To leading order in the HP expansion,
\begin{equation}
\hat{\text{S}}^-
=
\sum_{\boldsymbol{\mathfrak{i}}}
\hat{\text{s}}_{\boldsymbol{\mathfrak{i}}}^-
\simeq
(2s)^{1/2}
\sum_{\boldsymbol{\mathfrak{i}}}
\hat{\text{m}}_{\boldsymbol{\mathfrak{i}}}^{\dagger}
=
(2Ns)^{1/2}
\hat{m}_{\boldsymbol{0}}^{\dagger}.
\label{collective_uniform_equivalence}
\end{equation}
On the other hand, the collective HP transformation gives
\[
\hat{\text{S}}^-
\simeq
(2Ns)^{1/2}
\hat{\text{M}}^{\dagger}.
\]
Thus, in the low-excitation regime,
\begin{equation}
\hat{\text{M}}
\equiv
\hat{m}_{\boldsymbol{0}},
\label{collective_boson_uniform_mode}
\end{equation}
up to the notational distinction between the collective-spin and reciprocal-space descriptions. 
The operator \(\hat{\text{M}}\) therefore annihilates a spin flip delocalized over the entire ferromagnet, rather than a spin flip localized at a particular atomic site. 
This is precisely the uniform mode commonly retained in cavity-magnonics experiments.

The collective route therefore does not introduce a different physical excitation. 
Rather, it reaches the same uniform-mode description more directly by exploiting from the outset the fact that the experimentally relevant fields couple predominantly to the total spin. 
In particular, the nonlinear corrections of the uniform mode are obtained directly from the expansion of the collective HP transformation, without first generating the full set of interaction terms among reciprocal-space spin-wave modes.

When the in-plane field is approximately resonant with the magnon frequency, one can further simplify Eq.~\eqref{mag_ham_magneto_simp} by applying the rotating-wave approximation (RWA). 
Including the free Hamiltonians of the in-plane field, out-of-plane field, and magnon mode, and writing the magnon frequency as \(\omega_m\simeq g_0\), the resonant part of the Hamiltonian becomes
\begin{align}
    \hat{H}^{(\mathrm{RWA})}
    =&
    \hat{H}_0^{(\mathrm{RWA})}
    -
    Nsg_b
    \left(
    \hat{b}
    +
    \hat{b}^{\dagger}
    \right)
    \nonumber\\
    &
    +
    g_b
    \left(
    \hat{b}
    +
    \hat{b}^{\dagger}
    \right)
    \hat{\text{M}}^{\dagger}\hat{\text{M}}
    \nonumber\\
    &
    +
    \dfrac{g_c}{4(2Ns)^{1/2}}
    \left(
    \hat{\text{M}}^{\dagger 2}
    \hat{\text{M}}\hat{c}
    +
    \hat{\text{M}}^{\dagger}
    \hat{\text{M}}^2
    \hat{c}^{\dagger}
    \right),
    \label{RWA_ham}
\end{align}
where
\begin{align}
\hat{H}_0^{(\mathrm{RWA})}
=&
\omega_c
\hat{c}^{\dagger}\hat{c}
+
\omega_m
\hat{\text{M}}^{\dagger}\hat{\text{M}}
+
\omega_b
\hat{b}^{\dagger}\hat{b}
\nonumber\\
&
-
\dfrac{g_c}{2}(2Ns)^{1/2}
\left(
\hat{\text{M}}\hat{c}^{\dagger}
+
\hat{\text{M}}^{\dagger}\hat{c}
\right).
\label{RWA_free_ham}
\end{align}

The RWA keeps the excitation-conserving terms with respect to the in-plane field and neglects counter-rotating terms such as
\(\hat{\text{M}}\hat{c}\) and
\(\hat{\text{M}}^{\dagger}\hat{c}^{\dagger}\).
The out-of-plane field, in contrast, couples to the magnon number and acts as a modulation of the magnon frequency.

\section{Finite-spin correction to the magnon--photon coupling}

The collective-spin formulation also gives a direct physical interpretation of the leading nonlinear correction. 
Because the ferromagnet has a finite total spin \(S=Ns\), the uniform mode cannot be populated indefinitely as an ideal harmonic oscillator. 
This finite-spin character appears as a weak occupation dependence of the magnon--photon coupling.

To see this, consider the in-plane interaction before expanding the collective HP square root. 
In the rotating-wave approximation, the resonant part of the coupling may be written as
\begin{equation}
\hat H_{\mathrm{int}}^{\mathrm{RWA}}
=
-\frac{g_c}{2}
\left(
\hat c\,\hat S^-
+
\hat c^\dagger \hat S^+
\right).
\end{equation}
Let \(\hat n_M=\hat{\text{M}}^\dagger\hat{\text{M}}\). 
The matrix element coupling the states
\(\ket{n_M,1_c}\) and \(\ket{n_M+1,0_c}\), corresponding to the conversion of one cavity photon into one additional uniform magnon, is
\begin{equation}
G_{n_M}
=
\frac{g_c}{2}
(2Ns)^{1/2}
\sqrt{n_M+1}
\left(
1-\frac{n_M}{2Ns}
\right)^{1/2}.
\end{equation}
In the purely bosonic approximation, this matrix element would be
\begin{equation}
G_{n_M}^{(0)}
=
\frac{g_c}{2}
(2Ns)^{1/2}
\sqrt{n_M+1}.
\end{equation}
Therefore, the finite-spin correction is
\begin{equation}
\frac{G_{n_M}}{G_{n_M}^{(0)}}
=
\left(
1-\frac{n_M}{2Ns}
\right)^{1/2}
\simeq
1-\frac{n_M}{4Ns},
\qquad
n_M\ll 2Ns.
\end{equation}
Thus, the collective HP expansion predicts an occupation-dependent reduction of the effective magnon--photon coupling.

For a coherently driven uniform magnon mode with average occupation \(\bar n_M\), this result may be expressed as
\begin{equation}
G_{\mathrm{eff}}(\bar n_M)
\simeq
G_0
\left(
1-\frac{\bar n_M}{4Ns}
\right),
\qquad
G_0=\frac{g_c}{2}(2Ns)^{1/2}.
\end{equation}
The normal-mode splitting of the coupled magnon--photon system is therefore expected to decrease weakly as the uniform-mode population approaches a non-negligible fraction of the total spin length. 
This provides a direct physical consequence of the collective-spin derivation: the leading nonlinear correction describes the onset of macrospin saturation in the uniform mode. 
In weakly driven experiments, where \(\bar n_M\ll 2Ns\), this correction is negligible and the usual harmonic-oscillator description is recovered. 
Under strong in-plane driving, however, the correction gives a simple estimate of the regime in which finite-spin effects become observable.

To estimate the scale of this effect, we use the saturation magnetization of YIG,
\(M_s\simeq 140\,\mathrm{kA/m}\) ~\cite{prabhakar2009spin}.
For comparison, millimeter-sized YIG samples used in quantum
magnonics contain more than \(10^{19}\) spins~\cite{lachance2017resolving}. 
For a spherical sample of volume \(V\), the total spin length can be estimated from
\[
Ns \simeq \frac{M_s V}{g\mu_B},
\]
with \(g\simeq 2\). 
For a YIG sphere of diameter \(d\), this gives
\[
Ns
\simeq
3.95\times 10^{18}
\left(
\frac{d}{1\,\mathrm{mm}}
\right)^3
\left(
\frac{M_s}{140\,\mathrm{kA/m}}
\right).
\]
The fractional reduction of the coupling is therefore
\[
\frac{\Delta G}{G_0}
\simeq
\frac{\bar n_M}{4Ns}.
\]
A \(1\%\) reduction of the effective magnon--photon coupling would require
\[
\bar n_M \simeq 4\times 10^{-2}Ns.
\]
For a \(1\,\mathrm{mm}\)-diameter YIG sphere, this corresponds to
\(\bar n_M\simeq 1.6\times10^{17}\) uniform magnons, while for a
\(0.1\,\mathrm{mm}\)-diameter sphere it corresponds to
\(\bar n_M\simeq 1.6\times10^{14}\). 
At a magnon frequency of \(10\,\mathrm{GHz}\), these occupations correspond to
stored uniform-mode energies \(E_M\simeq \bar n_M\hbar\omega_m\) of order
\(10^{-6}\,\mathrm{J}\) and \(10^{-9}\,\mathrm{J}\), respectively, not including
losses or coupling inefficiencies associated with the external drive.
Thus, the finite-spin correction is negligible in the weak-drive quantum regime, but it may become relevant under strong coherent pumping, particularly for smaller YIG samples. 
This estimate also clarifies that the correction remains within the low-excitation regime, since even a \(1\%\) change corresponds to \(\bar n_M/(2Ns)\simeq 2\times10^{-2}\).

\section{\label{discussion}Discussion}

The conventional site-resolved Holstein--Primakoff construction and the collective-spin construction presented here lead to the same effective uniform-mode Hamiltonian in the long-wavelength regime relevant to cavity-magnonics experiments. 
The two approaches differ, however, in the intermediate variables used to reach this Hamiltonian and in the physical interpretation they make explicit. 
In the conventional route, the HP transformation is first applied to each atomic spin \(\hat{\boldsymbol{\text{s}}}_{\boldsymbol{\mathfrak{i}}}\), introducing local spin-flip operators \(\hat{\text{m}}_{\boldsymbol{\mathfrak{i}}}\). 
These local bosonic operators are then transformed to reciprocal space, giving magnon modes \(\hat{m}_{\boldsymbol{\lambda}}\). 
After this step, the long-wavelength character of the applied fields and the preparation near the fully polarized ground state are used to identify the uniform mode \(\boldsymbol{\lambda}=\boldsymbol{0}\) as the dominant degree of freedom coupled to the cavity fields.

The collective-spin route follows the opposite logic. 
Instead of first introducing local bosons and later projecting onto the uniform mode, one first identifies the uniform spin mode from the equations of motion. 
Since the applied fields are approximately uniform over the ferromagnetic sample and the system is prepared close to the fully polarized state, the relevant spin dynamics is carried by \(\hat{\boldsymbol{S}}_{\boldsymbol{0}}\). 
This mode is proportional to the total spin \(\hat{\textbf{S}}\), so that the effective Hamiltonian can be written directly in terms of collective spin operators. 
The HP transformation is then applied to \(\hat{\textbf{S}}\), introducing a single collective bosonic operator \(\hat{\text{M}}\) associated with spin flips of the entire ferromagnet.

The equivalence between the two approaches follows from the identification of the collective HP boson with the conventional uniform magnon operator. 
In the site-resolved treatment, the uniform magnon mode is the delocalized spin-flip operator
\[
\hat{m}_{\boldsymbol{0}}
=
\frac{1}{\sqrt{N}}
\sum_{\boldsymbol{\mathfrak{i}}}
\hat{\text{m}}_{\boldsymbol{\mathfrak{i}}}.
\]
In the low-excitation regime, the total spin lowering operator satisfies
\[
\hat{\text{S}}^-
=
\sum_{\boldsymbol{\mathfrak{i}}}
\hat{\text{s}}_{\boldsymbol{\mathfrak{i}}}^-
\simeq
(2s)^{1/2}
\sum_{\boldsymbol{\mathfrak{i}}}
\hat{\text{m}}_{\boldsymbol{\mathfrak{i}}}^{\dagger}
=
(2Ns)^{1/2}
\hat{m}_{\boldsymbol{0}}^{\dagger}.
\]
The collective HP transformation, on the other hand, gives
\[
\hat{\text{S}}^-
\simeq
(2Ns)^{1/2}
\hat{\text{M}}^{\dagger}.
\]
Therefore,
\[
\hat{\text{M}}\equiv \hat{m}_{\boldsymbol{0}},
\]
up to the notational distinction between the collective-spin and reciprocal-space descriptions. 
Thus, the collective-spin construction does not introduce a different excitation; it provides a direct route to the same uniform magnon mode.

The main conceptual difference is that the collective-spin method treats the ferromagnet, from the outset, as a single large spin of length \(Ns\) interacting with external fields. 
This makes explicit the macrospin interpretation underlying the uniform-mode approximation. 
In this sense, the ferromagnet behaves as a synthetic large-spin atom, or super-atom, whose collective spin can display magnonic analogues of phenomena familiar from atomic and molecular physics, such as Rabi oscillations, Autler--Townes splittings, and Floquet sidebands.

The collective formulation also gives a transparent perturbative hierarchy for nonlinear corrections. 
In the collective HP expansion, the relevant smallness condition is
\[
\braket{\hat{\text{M}}^{\dagger}\hat{\text{M}}}\ll 2Ns,
\]
which directly compares the number of collective excitations with the total spin length. 
In the conventional site-resolved formulation, the corresponding condition appears as a small local spin-flip density,
\[
\braket{
\hat{\text{m}}_{\boldsymbol{\mathfrak{i}}}^{\dagger}
\hat{\text{m}}_{\boldsymbol{\mathfrak{i}}}
}
\ll 2s.
\]
For a uniform delocalized excitation, this condition is equivalent to
\[
\sum_{\boldsymbol{\mathfrak{i}}}
\braket{
\hat{\text{m}}_{\boldsymbol{\mathfrak{i}}}^{\dagger}
\hat{\text{m}}_{\boldsymbol{\mathfrak{i}}}
}
=
\braket{
\hat{m}_{\boldsymbol{0}}^{\dagger}
\hat{m}_{\boldsymbol{0}}
}
\ll 2Ns.
\]
Thus, the two expansions are physically consistent, but the collective formulation expresses the relevant small parameter directly in terms of the total number of uniform-mode excitations.

Another useful feature of the collective approach is that the exchange interaction becomes a constant energy shift in the fully symmetric uniform sector. 
This follows from the fact that the total spin operators commute with the isotropic exchange interaction and that, in this sector, the total spin is fixed at \(\text{S}=Ns\). 
Consequently, the effective dynamics is governed only by the coupling of the collective spin to the static in-plane and out-of-plane fields. 
This provides a compact way of deriving the effective Hamiltonians relevant for driven cavity-magnonics experiments, especially when one is interested in nonlinear corrections associated with strong in-plane driving or Floquet modulation.

The usefulness of the collective-spin method is therefore not that it replaces the conventional spin-wave formalism in general. 
The site-resolved approach remains essential whenever nonuniform modes, finite-size effects, surface anisotropies, dipolar interactions, demagnetization fields, disorder, or multimode magnon dynamics are important. 
Rather, the collective approach is best understood as a direct and physically transparent formulation of the uniform-mode limit. 
Within this regime, it shortens the route from the microscopic Heisenberg Hamiltonian to the effective magnon--photon Hamiltonian and clarifies why the macroscopic ferromagnet can be treated as a synthetic large-spin object.

\section{\label{sec:conclusion}Conclusion}

We have presented a direct collective-spin derivation of the
effective uniform-mode Hamiltonian used in cavity magnonics.
Starting from the Heisenberg Hamiltonian for a ferromagnet coupled
to long-wavelength magnetic fields, we showed that, under the
conditions typically realized in cavity-magnonics experiments, the
relevant dynamics can be restricted to the uniform spin mode. In this
regime, the uniform mode is proportional to the total spin of the
ferromagnet, and the exchange interaction contributes only a constant
energy shift within the fully symmetric spin sector.

Applying the Holstein--Primakoff transformation directly to the total
spin provides a compact route to the usual bosonic description of the
uniform magnon mode. The resulting collective boson is equivalent to
the conventional uniform magnon operator obtained from the
site-resolved spin-wave formalism, but the collective-spin formulation
makes this identification explicit from the outset. It also expresses
the perturbative expansion in terms of the physically transparent
condition
\[
\langle \hat{\text{M}}^{\dagger}\hat{\text{M}}\rangle \ll 2Ns,
\]
which compares the number of collective spin flips with the total
spin length.

As a direct physical consequence, the leading nonlinear correction
produces an occupation-dependent reduction of the effective
magnon--photon coupling. For a coherently driven uniform mode with
average occupation \(\bar n_M\), the coupling is reduced according to
\[
G_{\mathrm{eff}}(\bar n_M)
\simeq
G_0
\left(
1-\frac{\bar n_M}{4Ns}
\right),
\]
within the low-excitation regime. This effect provides a simple
signature of finite-spin, or macrospin, saturation under strong
uniform-mode driving and suggests a possible power-dependent
correction to the normal-mode splitting in cavity-magnonics
experiments.

This perspective clarifies why a macroscopic ferromagnet in the
long-wavelength cavity regime can be treated as a synthetic
large-spin atom interacting with in-plane and out-of-plane
electromagnetic fields. The approach is not intended to replace the
full spin-wave formalism when nonuniform modes, dipolar effects,
demagnetization fields, or finite-size corrections are important.
Rather, it offers a streamlined and physically transparent formulation
of the uniform-mode limit central to many cavity-magnonics
experiments, while also identifying the leading finite-spin correction
that appears under strong driving.

\section*{Acknowledgments}

The authors acknowledge support from the São Paulo Research Foundation
(FAPESP, Grant No. 2024/08133-4), the National Council for Scientific
and Technological Development (CNPq, Grant No. 141561/2023-8), and the
University of Campinas. M.C.O. acknowledges partial financial support
from the National Institute of Science and Technology for Applied
Quantum Computing through CNPq Grant No. 408884/2024-0 and from
FAPESP through the Center for Research and Innovation on Smart and
Quantum Materials (CRISQuaM), Grant No. 2024/00998-6.

\appendix

\section{\label{eq_motion_spin}Equations of Motion of Nonuniform Spin Modes}

In this appendix, we give the equations of motion for the reciprocal-space spin modes when the applied fields are not restricted to the uniform component. 
In this more general case, a spin mode with wave vector \(\boldsymbol{\sigma}\) can couple to other modes through both the spatial dependence of the applied fields and the exchange interaction. 
This explicitly shows why the uniform-mode reduction used in the main text relies on the long-wavelength character of the fields and on the preparation of the ferromagnet close to the fully polarized state.

Using the Hamiltonian in Eq.~\eqref{full_spin_hamiltonian}, the commutation relations in Eqs. \eqref{comm_relations1} and \eqref{comm_relations}, and the centrosymmetry of the reciprocal-space summation, we obtain
\begin{align}
    \dfrac{d}{dt}\hat{S}_{\boldsymbol{\sigma}}^{-}
    =&
    +ig_0\hat{S}_{\boldsymbol{\sigma}}^-
    +
    ig_b
    \left(
    \hat{b}_{\boldsymbol{\lambda}_b}
    \hat{S}_{\boldsymbol{\sigma}+\boldsymbol{\lambda}_b}^-
    +
    \hat{b}_{\boldsymbol{\lambda}_b}^{\dagger}
    \hat{S}_{\boldsymbol{\sigma}-\boldsymbol{\lambda}_b}^-
    \right)
    \nonumber\\
    &
    -
    ig_c
    \left(
    \hat{c}_{\boldsymbol{\lambda}_c}
    \hat{S}_{\boldsymbol{\sigma}+\boldsymbol{\lambda}_c}^{z}
    +
    \hat{c}_{\boldsymbol{\lambda}_c}^{\dagger}
    \hat{S}_{\boldsymbol{\sigma}-\boldsymbol{\lambda}_c}^{z}
    \right)
    \nonumber\\
    &
    +
    2iJZ N^{-1/2}
    \sum_{\boldsymbol{\lambda}}
    \left(
    u_{\boldsymbol{\lambda}}
    -
    u_{\boldsymbol{\sigma}-\boldsymbol{\lambda}}
    \right)
    \hat{S}_{\boldsymbol{\sigma}-\boldsymbol{\lambda}}^{-}
    \hat{S}_{\boldsymbol{\lambda}}^{z},
    \label{desc_eq_motion}
\end{align}
\begin{align}
    \dfrac{d}{dt}\hat{S}_{\boldsymbol{\sigma}}^{+}
    =&
    -ig_0\hat{S}_{\boldsymbol{\sigma}}^+
    -
    ig_b
    \left(
    \hat{S}_{\boldsymbol{\sigma}+\boldsymbol{\lambda}_b}^{+}
    \hat{b}_{\boldsymbol{\lambda}_b}
    +
    \hat{b}_{\boldsymbol{\lambda}_b}^{\dagger}
    \hat{S}_{\boldsymbol{\sigma}-\boldsymbol{\lambda}_b}^{+}
    \right)
    \nonumber\\
    &
    +
    ig_c
    \left(
    \hat{c}_{\boldsymbol{\lambda}_c}
    \hat{S}_{\boldsymbol{\sigma}+\boldsymbol{\lambda}_c}^{z}
    +
    \hat{c}_{\boldsymbol{\lambda}_c}^{\dagger}
    \hat{S}_{\boldsymbol{\sigma}-\boldsymbol{\lambda}_c}^{z}
    \right)
    \nonumber\\
    &
    -
    2iJZ N^{-1/2}
    \sum_{\boldsymbol{\lambda}}
    \left(
    u_{\boldsymbol{\lambda}}
    -
    u_{\boldsymbol{\sigma}-\boldsymbol{\lambda}}
    \right)
    \hat{S}_{\boldsymbol{\sigma}-\boldsymbol{\lambda}}^{+}
    \hat{S}_{\boldsymbol{\lambda}}^{z},
    \label{asc_eq_motion}
\end{align}
and
\begin{align}
    \dfrac{d}{dt}\hat{S}_{\boldsymbol{\sigma}}^{z}
    =&
    \dfrac{ig_c}{2}
    \hat{c}_{\boldsymbol{\lambda}_c}
    \left(
    \hat{S}_{\boldsymbol{\sigma}+\boldsymbol{\lambda}_c}^{+}
    -
    \hat{S}_{\boldsymbol{\sigma}+\boldsymbol{\lambda}_c}^{-}
    \right)
    \nonumber\\
    &
    +
    \dfrac{ig_c}{2}
    \hat{c}_{\boldsymbol{\lambda}_c}^{\dagger}
    \left(
    \hat{S}_{\boldsymbol{\sigma}-\boldsymbol{\lambda}_c}^{+}
    -
    \hat{S}_{\boldsymbol{\sigma}-\boldsymbol{\lambda}_c}^{-}
    \right)
    \nonumber\\
    &
    +
    iJZ N^{-1/2}
    \sum_{\boldsymbol{\lambda}}
    \left(
    u_{\boldsymbol{\lambda}}
    -
    u_{\boldsymbol{\sigma}-\boldsymbol{\lambda}}
    \right)
    \hat{S}_{\boldsymbol{\sigma}-\boldsymbol{\lambda}}^{+}
    \hat{S}_{\boldsymbol{\lambda}}^{-}.
    \label{z_eq_motion}
\end{align}
For \(\boldsymbol{\sigma}=\boldsymbol{0}\) and uniform applied fields,
\[
\boldsymbol{\lambda}_b=\boldsymbol{\lambda}_c=\boldsymbol{0},
\]
the exchange terms vanish because
\[
u_{\boldsymbol{\lambda}}-u_{-\boldsymbol{\lambda}}=0,
\]
using the inversion symmetry of the nearest-neighbor lattice. 
The equations of motion then reduce to the closed uniform-mode equations used in the main text.

\bibliography{apssamp}

\end{document}